%% file: main.tex
\documentclass[sigconf]{acmart}
\usepackage{siunitx}
\usepackage{tabularx}
\usepackage{graphicx}
\usepackage{enumitem}
\usepackage{tikz}
\usepackage{placeins}
\usepackage{subcaption}
\usepackage{ifthen}
\newboolean{arxiv}
\setboolean{arxiv}{true} 

\AtBeginDocument{%
  }

\ifthenelse{\boolean{arxiv}}{
\settopmatter{printacmref=false, printccs=false, printfolios=false} 
\renewcommand\footnotetextcopyrightpermission[1]{}  
}
{
\setcopyright{acmlicensed}
\copyrightyear{2026}
\acmYear{2026}
\acmDOI{XXXXXXX.XXXXXXX}
\acmConference[SIGSPATIAL '26]{The 34th ACM International Conference on Advances in Geographic Information Systems}{Nov 03--06,
  2026}{Riverside, CA}
\acmISBN{978-1-4503-XXXX-X/2018/06}
}

\begin{document}


\ifthenelse{\boolean{arxiv}}{
\title{Simulating Public Transit Fare Policies in NYC: \\An Efficient, Socioeconomic-Aware Framework}
\renewcommand{\shorttitle}{Simulating Public Transit Fare Policies in NYC: An Efficient, Socioeconomic-Aware Framework}
}
{
\title{Simulating Public Transit Fare Policies in NYC: \\An Efficient, Socioeconomic-Aware Framework [Applications]}
\renewcommand{\shorttitle}{Simulating Public Transit Fare Policies in NYC: An Efficient, Socioeconomic-Aware Framework}
}


\input{meta/authors_short}

\begin{abstract}
Designing equitable and effective public transit fare policies is challenging due to complex interactions among traveler behavior, multimodal networks, and socioeconomic heterogeneity. This paper presents a scalable, data-driven simulation framework for evaluating transit fare policies in New York City (NYC), integrating a synthetic population, agent-based simulation, multimodal travel-time estimation, and fare-sensitive mode choice modeling. We evaluate multiple fare scenarios, including distance-based pricing, fare increases, and fare-free bus policies. Results show that pricing changes modestly affect total ridership but significantly alter modal composition and produce heterogeneous impacts across income groups. In particular, fare-free bus policies generate substantial benefits for lower-income riders by increasing bus usage and reducing fare burden, while introducing trade-offs in revenue. To support city-scale analysis, we introduce a sampling-based approach that reduces computational cost while preserving aggregate accuracy. The proposed framework provides a practical tool for assessing trade-offs between ridership, revenue, and equity, enabling more informed and equitable transit policy design.
\end{abstract}




\maketitle

\input{sections/introduction}
\input{sections/related_work_short}

\input{sections/data}

\input{sections/methods}

\input{sections/experiments}
\input{sections/conclusions}


\bibliographystyle{ACM-Reference-Format}
\bibliography{references}

\ifthenelse{\boolean{arxiv}}{
\clearpage
\appendix
\section*{Appendix}
\input{sections/calibration_parameters}
\input{sections/sampling_detail}

}{
}

\end{document}

%% file: meta/authors_short.tex
\settopmatter{authorsperrow=4} 

\author{Parker Wischhover}
\email{parker.wischhover@emory.edu}
\author{Hossein Amiri}
\orcid{0000-0003-0926-7679}
\email{hossein.amiri@emory.edu}

\affiliation{%
  \institution{Emory University}
  \city{Atlanta}
  \state{Georgia}
  \country{USA}
}

\author{Kiara Ha}
\email{kiara.ha@emory.edu}
\author{Jooyoung Yoo}
\orcid{0000-0002-5634-7367}
\email{jooyoung.yoo@emory.edu}
\affiliation{%
  \institution{Emory University}
  \city{Atlanta}
  \state{Georgia}
  \country{USA}
}

\author{Lina Li}
\email{lina.li@emory.edu}
\author{Andreas Z{\"u}fle}
\orcid{0000-0001-7001-4123}
\email{azufle@emory.edu}
\affiliation{%
  \institution{Emory University}
  \city{Atlanta}
  \state{Georgia}
  \country{USA}
}

\author{Joon-Seok Kim}
\email{joonseok.kim@emory.edu}
\orcid{0000-0001-9963-6698}
\affiliation{%
    \institution{Emory University}
    \city{Atlanta}
    \state{Georgia}
    \country{USA}
}

%% file: sections/introduction.tex
\section{Introduction}
\label{sec:introduction}

Public transit systems are essential to modern metropolitan areas, enabling mobility, supporting economic productivity, and promoting environmental sustainability. In dense cities such as New York City (NYC), public transportation serves as a critical lifeline for millions of residents across diverse socioeconomic groups. Factors such as affordability, accessibility, and service reliability strongly influence individual mobility decisions and shape broader urban outcomes, including congestion, emissions, and social equity. Among these, fare policy is a particularly influential lever, as even modest pricing changes can alter ridership patterns, modal choices, and access to opportunities for vulnerable populations \cite{kaddoura2015optimal,zhang2024optimal}.

Recent policy discussions in NYC, including proposals to expand fare subsidies and introduce fare-free services, underscore the urgency of designing effective and equitable pricing strategies \cite{css_fair_fares_2024,mta_budget_2026}. However, transit policy design remains inherently complex, as interventions interact with heterogeneous traveler preferences, multimodal transportation networks, and socioeconomic disparities \cite{kamel2021integrated,schlenther2025ridepooling}. These systems also operate under practical constraints such as capacity limits, operational costs, and financial considerations \cite{stuntz2018transit,ren2025distributional}. Consequently, predicting the impacts of fare policies is challenging, as pricing changes often lead to heterogeneous and context-dependent behavioral responses across populations \cite{kaddoura2015optimal,zhang2024optimal}.

Simulation-based approaches, particularly agent-based models (ABMs), provide a powerful framework for evaluating transportation policies by capturing individual behavior and multimodal interactions \cite{divasson2025agent,jing2024evaluating}. Large-scale implementations such as MATSim-NYC demonstrate the feasibility of such approaches for real-world urban systems \cite{chow2020multi}, while prior studies have used synthetic populations to evaluate congestion pricing and built environment policies in NYC \cite{he2021validated,he2020evaluation}. In parallel, data-driven approaches have been increasingly adopted to model travel behavior and improve prediction of mode choice by capturing complex relationships among travel time, cost, and socioeconomic characteristics \cite{zhao2020prediction,cheng2019applying}. Despite these advances, existing approaches often rely on simplified assumptions, focus on specific policy mechanisms, or fail to jointly capture multimodal dynamics and fine-grained population heterogeneity at city scale.

To address these limitations, this paper presents a scalable geospatial agent-based simulation framework for evaluating public transit fare policies in NYC. The proposed framework integrates behaviorally realistic decision models with a synthetic population and a multimodal transportation network, enabling detailed analysis of ridership, revenue, and equity outcomes under alternative fare scenarios. To support city-scale deployment, we introduce efficient travel time estimation and sampling strategies that significantly reduce computational cost while preserving accuracy.

Our study is guided by the following research questions:
\begin{itemize}[leftmargin=*,nosep]
    \item \textbf{RQ1:} How do alternative fare policies for NYC subways and buses impact ridership patterns and modal choices across the urban system?
    \item \textbf{RQ2:} What are the implications of these fare policies on transit revenue and the financial sustainability of the system?
    \item \textbf{RQ3:} How do different fare structures affect equity outcomes across socioeconomic and demographic groups?
    \item \textbf{RQ4:} How can geospatial simulation be used to analyze and compare fare policy impacts at city scale?
\end{itemize}

By addressing these questions, this work contributes a scalable and behaviorally grounded simulation framework for transit policy evaluation. The proposed approach enables policymakers to assess trade-offs between ridership, revenue, and equity in a controlled setting, supporting more informed and inclusive decision-making.

The remainder of the paper is organized as follows. Section~\ref{sec:related_work} reviews related literature. Section~\ref{sec:data} describes the study area and data sources. Section~\ref{sec:method} presents the proposed methodology. Section~\ref{sec:experiments} reports experimental results and analysis. Finally, Section~\ref{sec:conclusion} concludes the paper and outlines future work.

%% file: sections/related_work_short.tex
\vspace{-6pt}
\section{Related Work}
\label{sec:related_work}

Urban mobility modeling plays a central role in transportation planning and policy evaluation. Simulation-based approaches, particularly agent-based models (ABMs), are widely used to evaluate interventions such as congestion pricing and fare policy before real-world deployment \cite{divasson2025agent,chow2020multi}. In cities like NYC, large-scale frameworks with synthetic populations have been applied to study congestion pricing, built environment policies, and distributional impacts \cite{he2020evaluation,he2021validated,ren2025distributional}. At the same time, limited availability of high-resolution mobility data has motivated the use of synthetic populations and integrated data-driven approaches.

\vspace{-5pt}
\subsection{Agent-Based Mobility Simulation}

ABMs enable modeling of individual-level travel behavior and emergent system dynamics, making them well-suited for urban transportation systems \cite{divasson2025agent}. Platforms such as MATSim and SimMobility have been used extensively to evaluate policy interventions, including pricing strategies and infrastructure changes \cite{chow2020multi,jing2024evaluating,adnan2020examining}. Prior work shows that ABMs can capture behavioral responses to pricing, such as peak shifting and mode substitution \cite{lovric2016evaluating,jing2024evaluating}. 

However, most existing studies focus on specific interventions and often simplify interactions between multimodal systems and heterogeneous populations. This limits their ability to evaluate integrated fare policies at city scale.

\vspace{-5pt}
\subsection{Mode Choice Modeling}

Mode choice has been studied using both econometric and machine learning approaches. Discrete choice models, such as multinomial logit, provide interpretable representations of traveler decision-making based on time, cost, and other attributes \cite{kaddoura2015optimal,kamel2021integrated}. Machine learning methods can capture nonlinear patterns in behavior \cite{zhao2020prediction,cheng2019applying}, but typically sacrifice interpretability.

Recent work highlights the importance of incorporating socioeconomic heterogeneity into mode choice models, including income effects and intermodal trade-offs \cite{10.1145/3772318.3790772,schlenther2025ridepooling}. In particular, affordability plays a critical role in shaping travel decisions, especially for lower-income populations \cite{10838860}.

\vspace{-5pt}
\subsection{Multimodal Simulation and Scalability}

Multimodal simulation frameworks capture interactions across transportation modes and enable large-scale policy analysis. Systems such as MATSim-NYC and distributed simulation architectures demonstrate the feasibility of modeling millions of agents \cite{chow2020multi,10.1145/3764921.3770146}. 

Recent work has explored scalable optimization and simulation techniques, including data-driven and Bayesian approaches for policy evaluation \cite{11423433}. However, jointly modeling multimodal routing, large-scale populations, and computational efficiency remains challenging, particularly for repeated policy experiments.

\input{tabs/data_table}

\vspace{-5pt}
\subsection{Fare Policy and Equity Analysis}

Fare policy influences ridership, revenue, and equity outcomes. Prior studies examine pricing strategies such as marginal cost pricing and time-based fares \cite{kaddoura2015optimal,adnan2020examining}, as well as distributional impacts across income groups \cite{ren2025distributional,zhang2024optimal}. 

However, most work either focuses on pricing optimization or aggregate system effects. Fewer studies evaluate fare policies using behaviorally realistic, large-scale simulation frameworks that capture heterogeneous responses across demographic and spatial groups.

\vspace{-5pt}
\paragraph{Summary.}
Prior research has advanced ABMs, mode choice modeling, and multimodal simulation independently, but integrating these components into a unified, scalable framework for evaluating fare policy remains limited. In particular, there is a gap in modeling fare impacts with fine-grained spatial and socioeconomic detail at city scale.

%% file: tabs/data_table.tex
\begin{table*}[t]
\centering
\caption{Summary of datasets used in this study\vspace{-2pt}}\vspace{-6pt}
\small
\begin{tabular}{p{2.8cm} p{6.6cm} p{7.3cm}}
\hline
\textbf{Data Source} & \textbf{Description} & \textbf{Purpose} \\
\hline
Advan Foot-Traffic \& POI Data \cite{advan_patterns} 
& Aggregated device-based visitation patterns and $\sim$60,000 POIs across NYC 
& Approximate spatial activity distribution, infer travel destinations, and construct OD flows \\

Travel Surveys \cite{fhwa2022nhts,nycdot2024survey,rhts2011} 
& National and NYC travel survey data capturing trip behavior, mode choice, and trip frequencies 
& Calibrate and validate behavioral models, mode choice, and travel demand patterns \\

NYC TLC High-Volume FHV Trip Records \cite{nyctlc_fhv}
& Ride-hail (for-hire vehicle) trip fares and distances, one month of records
& Fit the ride-hail base and per-km fare \\

LODES Workplace Area Characteristics \cite{lodes_wac}
& LEHD workplace employment counts (jobs per census block) for NYC, 2023
& Assign agent work locations (CBG) via job-weighted sampling \\

ATUS Data \cite{atus2023} 
& National time-use data capturing daily activity schedules and work patterns 
& Assign activity schedules, work participation, and departure times \\

ACS + PUMS Data \cite{acs2022} 
& Demographic and socioeconomic microdata from the U.S. Census Bureau 
& Input for synthetic population generation \\


OpenStreetMap \cite{openstreetmap2024} 
& Street network and pedestrian infrastructure data 
& Construct road and walking network for multimodal routing \\

GTFS Transit Data \cite{gtfs2023} 
& Public transit routes, schedules, and stop locations 
& Model transit network structure, service frequency, and routing \\

MTA Data \cite{mta2023} 
& Observed ridership and operational statistics 
& Calibrate and validate simulation outputs against real-world data \\
\hline
\end{tabular}\vspace{-6pt}
\label{tab:data}
\vspace{-6pt}
\end{table*}

%% file: sections/data.tex
\vspace{-6pt}
\section{Data and Study Area}
\label{sec:data}

\subsection{Study Area: New York City}

Our study focuses on NYC, one of the largest and most complex public transit systems in the world. The Metropolitan Transportation Authority (MTA) operates an extensive multimodal network of subways and buses that supports millions of daily trips across a dense and socioeconomically diverse urban environment. NYC’s high reliance on public transit, combined with pronounced spatial and demographic heterogeneity, makes it an ideal setting for evaluating fare policy impacts.

Fare policy in NYC has evolved substantially over time. As of January 2026, the system employs a flat fare regardless of distance or travel time. The transition from the legacy MetroCard system to OMNY, a contactless payment platform, has introduced fare capping and replaced traditional unlimited-ride passes \cite{mta_metrocard_sunset_2025}, fundamentally altering how frequent riders experience pricing.

Financial sustainability remains a major challenge. The MTA faces a persistent structural deficit and relies heavily on subsidies and dedicated taxes, with farebox revenue covering only a portion of operating costs \cite{mta_budget_2026,osc_dinapoli_mta_2025}. This highlights the tension between maintaining affordability and ensuring long-term system viability.

Equity considerations further complicate fare policy design. Transit usage is disproportionately higher among lower-income residents, who also face greater affordability challenges. Surveys indicate that roughly 30\% of low-income New Yorkers report difficulty paying for transit even with existing subsidy programs \cite{css_fair_fares_2024}. Programs such as Fair Fares provide important support but do not fully cover all eligible or near-eligible populations \cite{css_fair_fares_2024,mta_fair_fares_policy_2026}.

In response, policymakers have proposed alternative fare structures to improve accessibility and equity, including expanded subsidies and fare-free transit initiatives. For example, proposals for free bus service have gained prominence in recent discussions \cite{cbs_mamdani_free_buses_2023}, though they raise questions about financial feasibility and long-term sustainability.

Evaluating such policies through real-world implementation is costly and risky. NYC therefore provides a compelling setting for simulation-based analysis, enabling systematic assessment of fare policy impacts on ridership, revenue, and equity under realistic behavioral and network conditions.

\vspace{-12pt}
\subsection{Data}

To support a high-fidelity, city-scale simulation, we integrate multiple heterogeneous datasets capturing mobility behavior, transportation infrastructure, and population characteristics. Table~\ref{tab:data} summarizes the datasets used in this study and their respective roles in the framework.

Broadly, the data can be categorized into three groups: (1) \textit{mobility data}, (2) \textit{population and behavioral data}, and (3) \textit{transportation network data}. First, mobility patterns and activity locations are derived from Advan foot-traffic and point-of-interest (POI) data, which provide large-scale observations of visitation flows across approximately 60,000 locations in NYC. These data are used to calibrate spatial activity distributions and infer realistic travel destinations. 
Second, behavioral and demographic characteristics are informed by multiple sources. Travel surveys (national and NYC-specific) provide empirical estimates of trip-making behavior, mode choice, and travel frequencies. Additionally, the American Time Use Survey (ATUS) is used to inform distributions of daily activity patterns, including work schedules and departure times. Demographic microdata from the American Community Survey (ACS) and Public Use Microdata Sample (PUMS) are used as inputs to the Likeness framework \cite{tuccillo2022likeness} to generate a synthetic population of approximately 3.4 million agents, each with 13 demographic attributes. This enables a realistic and heterogeneous representation of NYC residents.
Third, transportation infrastructure is modeled using OpenStreetMap (OSM) for the street and pedestrian network and General Transit Feed Specification (GTFS) data for transit routes, schedules, and stops.
In addition, MTA data, including observed ridership and operational statistics, are used for calibration and validation to ensure consistency between simulated and real-world system behavior. 

These datasets enable a comprehensive, data-driven representation of NYC's multimodal transportation system and its heterogeneous population, forming the foundation for evaluating alternative fare policies.

\vspace{-10pt}
\subsection{Data Processing}

A key step in our framework is constructing an origin-destination (OD) matrix representing aggregate travel demand across NYC. Instead of relying on external OD data, we approximate this matrix from observed mobility patterns using Advan weekly foot-traffic data.
The dataset provides anonymized visitation counts to POIs along with inferred home locations at a zonal level. We aggregate these data into analysis zones aligned with our simulation geography, treating visits to POIs as proxy destinations and inferred home locations as origins. 

The OD matrix is constructed by (1) mapping POIs to zones, (2) assigning each visit to an origin zone using inferred home locations, and (3) aggregating weekly visit counts into zone-to-zone flows. These flows are then normalized and disaggregated to daily levels using scaling factors derived from travel survey data.

The resulting OD matrix serves as an aggregate representation of travel demand for calibration and validation of the agent-based simulation. By leveraging observed mobility patterns, this approach preserves spatial heterogeneity and improves the representation of trip distribution and destination choice across NYC.

%% file: sections/methods.tex
\vspace{-8pt}
\section{Methods}
\label{sec:method}

\begin{figure}[h]
    \centering
    \includegraphics[width=\linewidth]{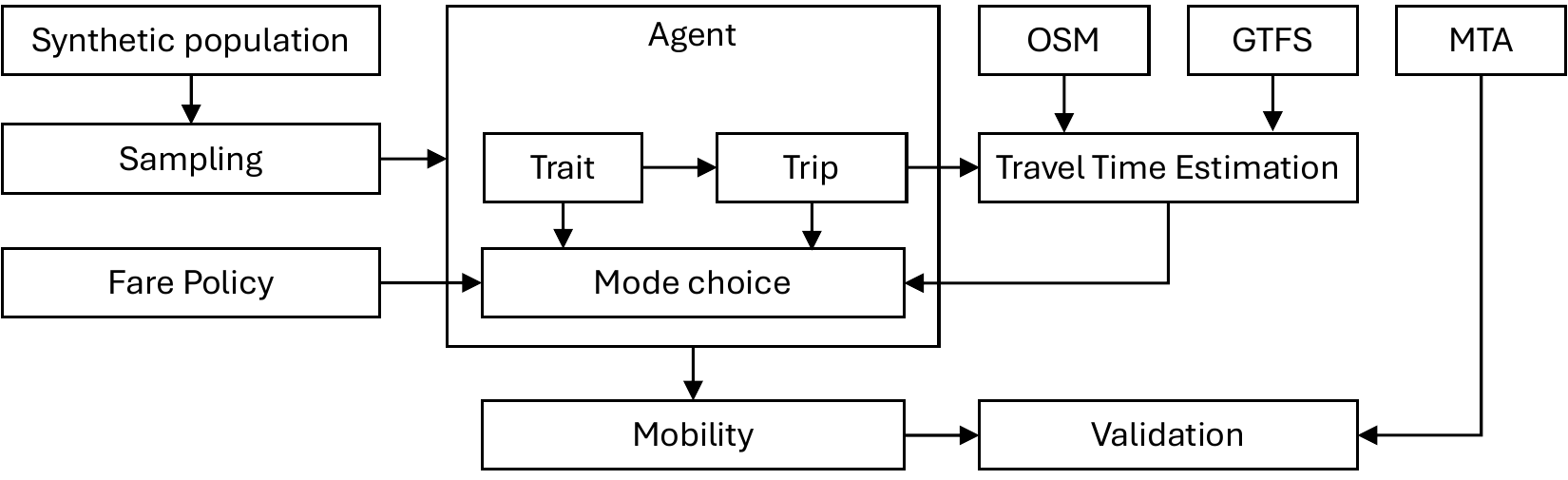}\vspace{-6pt}
    \caption{Workflow of the proposed simulation pipeline, from population sampling to trip generation, mode choice, and validation.\vspace{-5pt}}\vspace{-6pt}
    \label{fig:workflow}
\end{figure}

This section describes a simulation framework for evaluating public transit fare policies in NYC. The framework combines (i) agent-based behavioral modeling, (ii) multimodal travel-time estimation, (iii) cost-efficient sampling, and (iv) data-driven calibration into a unified pipeline (Figure~\ref{fig:workflow}).

Starting from a synthetic population, agents are sampled and assigned demographic and behavioral attributes that drive activity scheduling and trip generation. Travel costs are computed using a multimodal travel-time module based on OSM and GTFS data, capturing both walking and transit options. Mode choices are then determined using cost- and fare-sensitive decision models under different policy scenarios. Finally, simulated trips are aggregated and validated against observed MTA and mobility datasets to ensure consistency with real-world patterns.

\vspace{-6pt}
\subsection{Agent-Based Mobility Simulation}
\label{sec:agent_logic}

\begin{figure}
    \centering
    \includegraphics[width=1\linewidth]{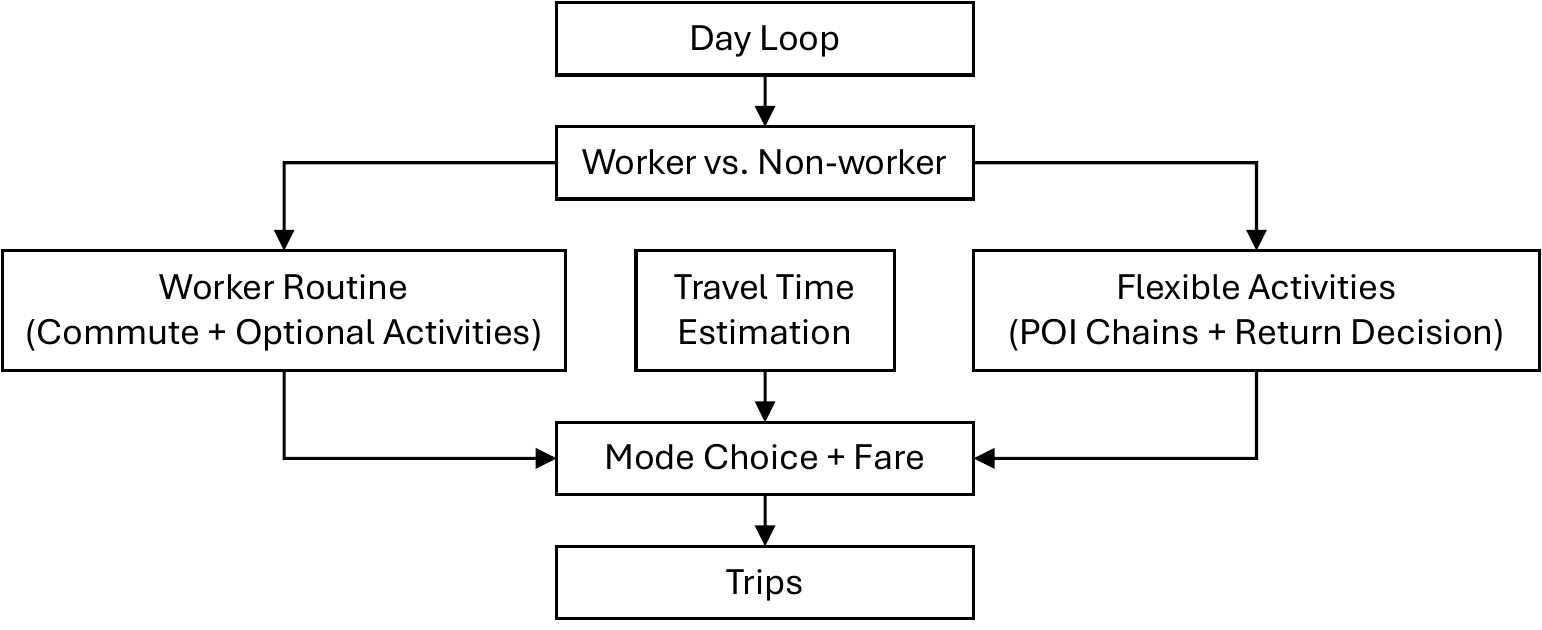}\vspace{-6pt}
    \caption{Simplified agent-level mobility simulation logic, showing activity patterns and sequential trip and mode choice decisions under travel-time and fare constraints.\vspace{-5pt}}\vspace{-8pt}
    \label{fig:agent_logic}
    
\end{figure}

This subsection describes how individual travel behavior is simulated, including activity scheduling, trip generation, and travel decisions. We simulate mobility at the individual level, where each agent generates a sequence of trips over a one-week horizon to capture weekday–weekend behavioral variation while maintaining tractability. The model captures heterogeneity in demographics, preferences, and travel behavior, while maintaining scalability for millions of agents.

\vspace{-4pt}
\subsubsection{Behavioral Structure}
Agents follow daily schedules that depend on employment status. Workers exhibit structured weekday patterns consisting of home-work commutes, optional lunch sub-tours, and discretionary post-work activities. Non-workers and weekends follow more flexible patterns, where trips are generated through stochastic activity chains with probabilistic stopping and return-home behavior. Figure~\ref{fig:agent_logic} illustrates this process.

\vspace{-4pt}
\subsubsection{Trip Generation}
Trips are generated sequentially based on the agent’s current location and time. Destinations are sampled from POIs using a probabilistic destination choice model based on distance decay and activity-type preferences, and agent-specific attributes (e.g., distance aversion, income sensitivity). Time is updated dynamically through travel and dwell durations.

\vspace{-4pt}
\subsubsection{Travel and Mode Decisions}
For each generated trip, travel time is obtained from the precomputed
multimodal matrices for the relevant mode and
time period. The traveler then selects a mode from a discrete-choice
behavioral model in which fare policy enters the utility of transit
alternatives directly, normalized by traveler income, so that the same fare
imposes a larger disutility on lower-income agents. The mode is sampled from
the resulting choice probabilities.

\vspace{-4pt}
\subsubsection{Simulation Output}
The simulation produces a sequence of trips for each agent, including origin, destination, departure time, travel duration, mode, and fare. These trajectories are aggregated to derive system-level metrics such as ridership patterns, travel demand, and accessibility.

\vspace{-6pt}
\subsection{Scalable Multimodal Travel Time Estimation}
\label{sec:travel_time}

This subsection presents a scalable method for estimating multimodal travel times used in destination choice and mode selection. Efficient and scalable estimation of travel time is a critical component of large-scale agent-based simulation frameworks, as it directly affects both destination choice and mode selection \cite{rasouli2014activity, w2016multi}. To support simulation across millions of individuals, we precompute a travel time matrix between CBG centroids. This approach significantly reduces runtime costs compared to on-demand routing while preserving sufficient spatial resolution for policy analysis \cite{bast2016route}.

\vspace{-4pt}
\subsubsection{Problem Formulation}
We estimate travel times as a function of origin CBG, destination CBG, departure time (or time period), and transportation mode. The model supports multiple modalities, including walking, biking, driving, and public transit (subway and bus). Formally, the travel time function is defined as: $T(o, d, t, m)$,
where $o$ and $d$ are origin and destination CBGs, $t$ denotes departure time (or time window), and $m$ is the transportation mode.

\vspace{-4pt}
\subsubsection{Landmark-Based Decomposition.}
Direct computation of all pairwise CBG-to-CBG travel times is computationally expensive due to the large number of zones. To address this, we adopt a landmark-based approximation strategy inspired by hub-labeling and landmark routing methods \cite{abraham2011hub, phan2019fastpublictransitrouting}. A set of representative locations, i.e., 500 landmarks empirically chosen to balance approximation accuracy and preprocessing cost, is selected across NYC, and travel times are computed between each landmark and all CBG centroids. Pairwise CBG-to-CBG travel times are then approximated via these intermediates as:
\[
T(o,d) \approx \min_{l \in L} \left( T(o,l) + T(l,d) \right).
\]

This design reduces the number of required shortest-path computations and enables scalable preprocessing for large urban networks while maintaining good approximation quality \cite{bast2016route}.

\vspace{-4pt}
\subsubsection{Road Network-Based Routing (Walk, Bike, Drive).}
For walking, biking, and driving modes, we use OSM data processed via OSMnx to construct detailed street and pedestrian networks \cite{boeing2017osmnx}. For each mode, we perform single-source Dijkstra shortest-path computations from each landmark node using NetworkX \cite{hagberg2008exploring}.
For walking and biking, the graph is effectively undirected, allowing reuse of landmark-to-CBG shortest path results by transposition. For driving, directionality is critical due to one-way streets; therefore, we perform an additional routing pass on the reversed graph to obtain accurate CBG-to-landmark travel times.
This approach reduces the number of shortest-path computations substantially (e.g., from thousands of CBG origins to 500 landmark runs), enabling scalable computation across the city.

\vspace{-4pt}
\subsubsection{Transit Routing (Subway and Bus).}
For public transit modes, travel time estimation is time-dependent due to schedules, transfer constraints, and waiting times. We use the r5py routing engine with GTFS data and OSM networks to compute multimodal transit travel times \cite{bast2016route, phan2019fastpublictransitrouting}.
Unlike road-based routing, transit travel times cannot be inferred via simple matrix transposition because departure-time-dependent factors differ by direction. Therefore, we compute both landmark-to-CBG and CBG-to-landmark travel time matrices explicitly for each mode and time period.
To capture temporal variability, we define multiple representative time windows (e.g., weekday AM peak, midday, PM peak, nighttime, and weekend). A separate travel time matrix is generated for each mode and time period combination.

\vspace{-4pt}
\subsubsection{Output Representation.}
The resulting travel time matrices are stored in a compact format for each mode and (for transit) each time period. Each matrix captures travel times (in seconds) between CBGs and landmarks in both directions. During simulation, travel time between any origin-destination pair is approximated using these precomputed matrices.

\vspace{-5pt}
\subsection{Fare-Sensitive Mode Choice}
\label{sec:mode_choice}

Given a generated trip, an agent chooses among competing travel modes. We model
this with a multinomial logit (MNL) discrete-choice model, which links each
mode's monetary cost, travel time, and the traveler's socioeconomic attributes
to a choice probability. We adopt a discrete-choice formulation rather than a
black-box classifier because it makes fare effects explicit and interpretable:
the fare enters the utility of transit alternatives directly and is normalized
by traveler income, which is the mechanism through which pricing changes
propagate into modal shifts. This
property is essential for equity analysis.

\paragraph{Choice Set and Utility.}
An agent chooses among up to six modes---subway (the reference alternative),
bus, car, walk, bike, and ridehail. Each mode $m$ is assigned a utility built
from an alternative-specific constant $\alpha_m$, the trip's travel time
$\tau_m$, its income-normalized cost burden $b_m$ (defined below), and a few
mode-specific terms $\gamma_m z_m$:
\begin{equation}
U_m \;=\; \alpha_m \;+\; \beta_t\,\tau_m \;+\; \beta_c\,b_m \;+\; \gamma_m z_m .
\label{eq:utility}
\end{equation}
The travel-time and cost coefficients $\beta_t,\beta_c$ are shared across modes,
while $\gamma_m z_m$ gathers covariates that matter only for particular modes:
peak-period timing for transit, income and trip purpose for car, and traveler
age for walk and bike (full list in Table~\ref{tab:mnl}). The agent then selects
each available mode with the standard logit probability
\begin{equation}
P(m) \;=\; \frac{\exp(U_m)}{\sum_{k}\exp(U_k)} ,
\label{eq:logit}
\end{equation}
where the sum runs only over modes available for the trip: walk and bike for
short trips (under $2$ and $8$\,km), car only with household vehicle access, and
transit and ridehail throughout the network.

\paragraph{Income-Normalized Fare Sensitivity.}
The cost term is the policy-sensitive core of the model. We express monetary
cost as a \emph{cost burden}, the trip's out-of-pocket cost relative to the
agent's hourly income $w$,
\begin{equation}
b_m \;=\; \frac{c_m}{w},
\label{eq:burden}
\end{equation}
where $c_m$ is the out-of-pocket cost of taking mode $m$ for the trip: the fare
for transit (set by the active fare policy), a distance- and borough-dependent
operating-plus-parking cost for car, a base-plus-distance fare for ridehail, and
zero for walk and bike. A single coefficient $\beta_c$ scales this burden across
all modes. Because
$b_m$ divides cost by income, a fixed fare increase raises the disutility of
transit more for low-income travelers than for high-income travelers, yielding
income-differentiated fare elasticities without any group-specific parameters.
All monetary costs and incomes are expressed on a single (current-dollar) basis
so that $\beta_c$ is estimated and applied at the same scale.

\paragraph{Estimation.}
Coefficients are estimated by maximum likelihood on revealed-preference trips
from the NYC Regional Household Travel Survey (RHTS)~\cite{rhts2011}, the most
recent large-scale household travel diary with mode and borough detail for the
region ($n=\num{29613}$ trips after filtering to NYC residents). Each
observation contributes its survey expansion weight, normalized to unit mean. We
optimize the weighted log-likelihood with L-BFGS-B under sign constraints
($\beta_t\!\le\!0$, $\beta_c\!\le\!0$). A
held-out $L_2$ penalty sweep selected no regularization, indicating the
specification is not overparameterized. Modal travel times for estimation are
drawn from the same landmark-based matrices used in simulation
(Section~\ref{sec:travel_time}). We find that replacing distance/speed proxies
with these network-derived travel times is associated with an identifiable improvement in fit (in-sample pseudo-$R^2$
rising from $0.24$ to $0.46$).

\begin{table}[!ht]
\centering
\vspace{-5pt}
\caption{Estimated MNL coefficients (n=29,613)}\vspace{-5pt}
\label{tab:mnl}
\small
\begin{tabular}{llr}
\toprule
Group & Parameter & Estimate \\
\midrule
Travel time     & $\beta_t$ (per min)              & $-0.010$ \\
Cost burden     & $\beta_c$                         & $-1.962$ \\
Transit peak    & $\beta_{\text{AM}}$ / $\beta_{\text{PM}}$ & $0.455$ / $0.336$ \\
Car             & income tier                       & $-0.259$ \\
                & non-home-based                    & $0.806$ \\
Walk / Bike     & age                               & $-0.002$ / $0.015$ \\
\midrule
ASC             & bus / car / walk / bike           & $-0.31$ / $2.66$ / $2.64$ / $-2.52$ \\
\bottomrule
\end{tabular}\vspace{-8pt}
\end{table}

\paragraph{Fit and Validation.}
On the full estimation sample the model attains a McFadden pseudo-$R^2$ of
$0.46$, and its coefficients carry the expected signs (Table~\ref{tab:mnl}):
travel time and cost burden reduce utility, peak-period indicators raise transit
utility, and the non-home-based term raises car utility. Because an in-sample
fit is optimistic, we treat held-out performance as the primary evidence of
generalization, assessing it with stratified $k$-fold cross-validation and
sampling a predicted mode per trip from~\eqref{eq:logit} to mirror the
simulation's stochastic choice rule rather than a deterministic argmax. The
held-out pseudo-$R^2$ falls to $0.21$ but remains above the $0.2$ value
conventionally regarded as a good fit for discrete-choice models. The gap from
the in-sample figure reflects this optimism rather than over-parameterization,
as the $13$-parameter specification is estimated on $\num{29613}$ observations
and the $L_2$ penalty sweep selected no regularization. The structural component
of the fit generalizes well: across folds the model reproduces the aggregate
RHTS modal split to within $1.4$ percentage points for every surveyed mode at a
$59.3\%$ hit rate (Table~\ref{tab:mnl-validation}). As an independent check, the
cross-validated shares are also broadly consistent with an external NYC DOT
mode-share survey~\cite{nycdot2024survey} not used in estimation (predicted vs.\
surveyed: walk $37.5$ vs.\ $37.6\%$, car $34.5$ vs.\ $38.5\%$, subway $16.0$
vs.\ $13.8\%$, bus $10.2$ vs.\ $7.2\%$, bike $1.8$ vs.\ $2.9\%$), corroborating
that the behavioral model recovers realistic mode shares before it is embedded
in the city-scale simulation.

\begin{table}[!h]
\centering
\vspace{-5pt}
\caption{Cross-validated mode-share recovery: observed (RHTS) vs.\ predicted
(sampled) shares, mean over 5 folds}\vspace{-5pt}
\label{tab:mnl-validation}
\small
\begin{tabular}{lrr}
\toprule
Mode & Observed & Predicted \\
\midrule
Subway   & $15.3\%$ & $16.0\%$ \\
Bus      & $9.8\%$  & $10.2\%$ \\
Car      & $35.9\%$ & $34.5\%$ \\
Walk     & $37.1\%$ & $37.5\%$ \\
Bike     & $1.8\%$  & $1.8\%$ \\
\bottomrule
\end{tabular}\vspace{-10pt}
\end{table}

\vspace{-5pt}
\subsection{Scalable Sampling Framework}
\label{sec:sampling}

To reduce computational cost while preserving accuracy, we adopt a two-stage sampling framework tailored to large-scale simulation. Running the full synthetic population of \num{3409887} agents for every candidate fare policy is computationally expensive and limits the ability to explore multiple scenarios. The proposed approach enables efficient experimentation while maintaining consistency with full-population results.

In the first stage, CBGs are characterized using aggregated population attributes and grouped into clusters based on socioeconomic, spatial, and behavioral features. Representative CBGs are then selected from each cluster to capture the diversity of urban travel patterns. This stage is primarily used to reduce the computational cost of validation experiments while preserving heterogeneity across neighborhoods.



In the second stage, agents are sampled within each selected CBG using either simple random sampling (SRS) or stratified sampling designs. The purpose of this stage is to determine how much of the synthetic population must be simulated to recover key full-population outcomes with acceptable accuracy. Sampling is performed without replacement, and sampled agents are assigned inverse-probability weights so that aggregate outcomes approximate the corresponding 100\% benchmark simulation.

We compare SRS with stratified designs based on income, spatial location, and commute-time characteristics. For each sampling strategy and sampling rate, the simulation is repeated multiple times (20 times) and compared with the 100\% benchmark over a set of target metrics, including transit ridership, fare revenue, mode-share error, and demographic composition measures. Performance is summarized using the mean relative error across the target metrics, with uncertainty represented by the standard deviation across repeated samples.

This framework enables systematic evaluation of trade-offs between computational efficiency and accuracy. In particular, it allows us to determine the minimum sampling rate required to maintain reliable estimates, thereby supporting scalable analysis of multiple fare policy scenarios. Detailed descriptions of feature construction, clustering, sampling strategies, and evaluation criteria are provided in 
\ifthenelse{\boolean{arxiv}}{
Appendix~\ref{sec:appendix_sampling}.
}
{
\cite{wischhover26simulating_arxiv}.
}

\subsection{Calibration}
\label{sec:calibration}

We calibrate model parameters to align simulated mobility patterns with observed data. Calibration of large-scale agent-based models is challenging due to high-dimensional parameter spaces and nonlinear interactions between behavior and network dynamics. Following prior work on scalable mobility simulation systems \cite{amiri2025hd,amiri2026hd}, we adopt a data-driven approach that leverages aggregated simulation outputs and empirical observations.

We formulate calibration as an optimization problem:
\[
\min_\theta \sum_k w_k \| y_k^{\text{sim}}(\theta) - y_k^{\text{obs}} \|,
\]
where $\theta$ represents model parameters and $y_k$ denotes multiple target metrics. Calibration targets include aggregate ridership, spatial trip distributions, and temporal activity patterns derived from Advan mobility data and MTA statistics. This multi-objective formulation ensures that both system-level trends and spatial heterogeneity are captured.

We optimize this objective using a genetic algorithm (GA), which is well-suited for exploring high-dimensional, non-convex spaces without requiring gradient information. Candidate parameter sets are iteratively evolved through selection, crossover, and mutation, with fitness evaluated based on agreement with observed data.
To improve computational efficiency, calibration is performed using aggregated simulation outputs rather than full agent trajectories. Simulation runs are parallelized across compute nodes, enabling scalable evaluation of candidate parameter sets. 
The calibrated parameters govern key behavioral components, including destination choice, activity scheduling, and cost sensitivity. Detailed parameter definitions, search ranges, and constraints are provided in 
\ifthenelse{\boolean{arxiv}}{
Appendix~\ref{sec:appendix_calibration}. 
}
{
\cite{wischhover26simulating_arxiv}.
}
The resulting parameter sets produce simulations that reproduce observed mobility patterns and serve as the basis for evaluating fare policy impacts on ridership, revenue, and equity outcomes.

%% file: sections/experiments.tex
\begin{figure*}
    \centering
    \includegraphics[width=0.85\linewidth]{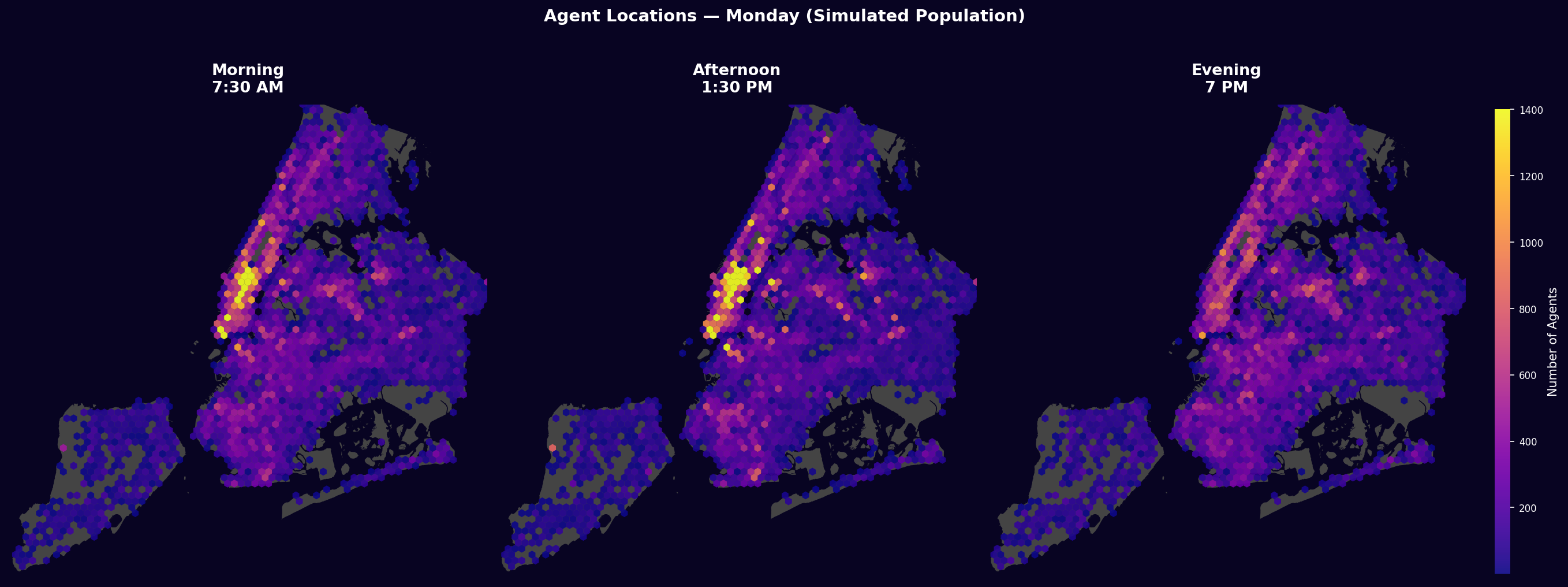}\vspace{-8pt}
    \caption{Weekday spatial and temporal patterns of simulated travel demand across NYC.\vspace{-8pt}}
    \label{fig:weekday_heatmap}
\end{figure*}

\section{Experiments}
\label{sec:experiments}

This section evaluates the proposed framework with respect to the research questions outlined in Section~\ref{sec:introduction}. Specifically, we assess (RQ1) ridership and mode choice impacts, (RQ2) revenue implications, (RQ3) equity outcomes, and (RQ4) the effectiveness of geospatial simulation for policy analysis. 
We evaluate the framework along three dimensions: (i) behavioral realism, (ii) system-level outcomes under alternative fare policies, and (iii) computational efficiency. Unless otherwise specified, experiments are conducted on a MacBook M2 with 16GB memory, using a 10\% sampled synthetic population (340{,}979 agents) simulated over a one-week horizon.

\subsection{Experimental Setup}

We use the NYC synthetic population described in Section~\ref{sec:data}, consisting of approximately 3.4 million agents. The transportation network is constructed from OSM and GTFS data, with precomputed travel time matrices for all modes and time periods (Section~\ref{sec:travel_time}). 

All experiments are calibrated using the procedure described in Section~\ref{sec:calibration}, with parameters optimized to match observed mobility patterns derived from Advan and MTA data. We evaluate policies under the baseline NYC flat-fare structure and a set of alternative fare scenarios.
\begin{itemize}[nosep]
    \item \textbf{Baseline:} Flat fare (\$3.00)
    \item \textbf{Scenario A:} Distance-based fare (base \$1.50 + \$0.25/km)
    \item \textbf{Scenario B:} Raised flat fare (\$3.50)
    \item \textbf{Scenario C:} Fare-free buses
\end{itemize}

Evaluation focuses on three primary outcomes: ridership, revenue, and equity.

\subsection{Behavioral Validation}


\subsubsection{Spatiotemporal Distribution of Simulated Mobility}

Figure~\ref{fig:weekday_heatmap} shows the spatial distribution of simulated agent locations across NYC at three weekday time periods: morning (7:30 AM), afternoon (1:30 PM), and evening (7:00 PM), using hexagonal aggregation where color intensity indicates agent density.

Morning activity is concentrated in Manhattan and along major commuting corridors, reflecting inbound travel. By the afternoon, activity becomes more dispersed, with sustained density in central business districts and increased presence in mixed-use areas. In the evening, density shifts back toward residential areas in the outer boroughs, consistent with return-home patterns.

Overall, the simulation captures key spatiotemporal dynamics, including peak-hour concentration and daily redistribution, providing qualitative validation of realistic urban mobility patterns.


\begin{figure}
    \begin{subfigure}{0.49\textwidth}\vspace{-5pt}
        \centering
        \includegraphics[width=0.8\textwidth]{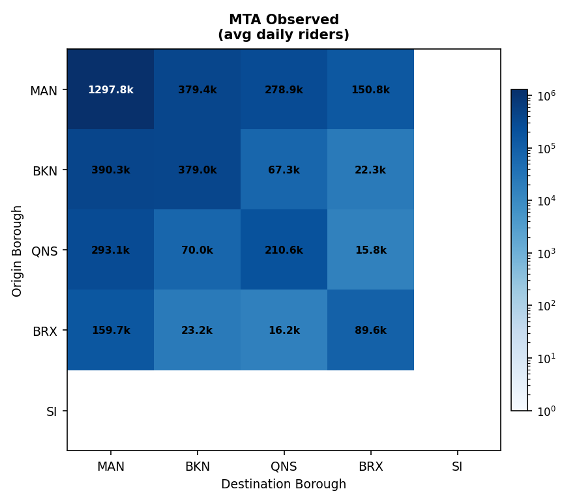}\vspace{-7pt}
        \caption{MTA observed}
        \label{fig:od_comparison_mta}
    \end{subfigure}
    \begin{subfigure}{0.49\textwidth}
        \centering
        \includegraphics[width=.8\textwidth]{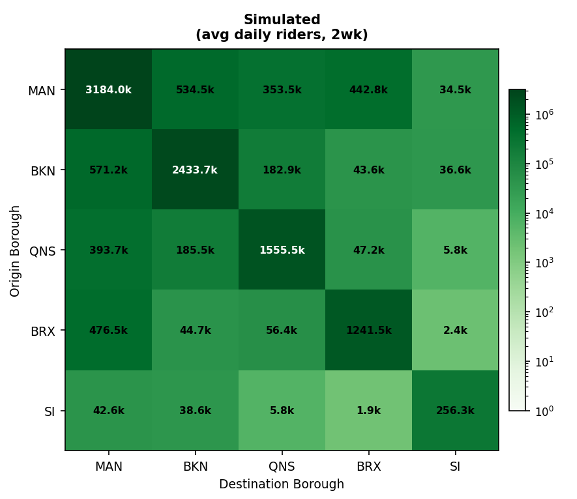}\vspace{-7pt}
        \caption{Simulated baseline}\vspace{-5pt}
        \label{fig:od_comparison_baseline}
    \end{subfigure}
    \caption{Borough-Pair Subway OD on Monday}\vspace{-12pt}
    \label{fig:od_comparison}
\end{figure}

\vspace{-5pt}

\begin{figure*}[h]
    \centering
    \begin{subfigure}{0.49\textwidth}\vspace{-6pt}
        \centering
        \includegraphics[width=0.95\textwidth]{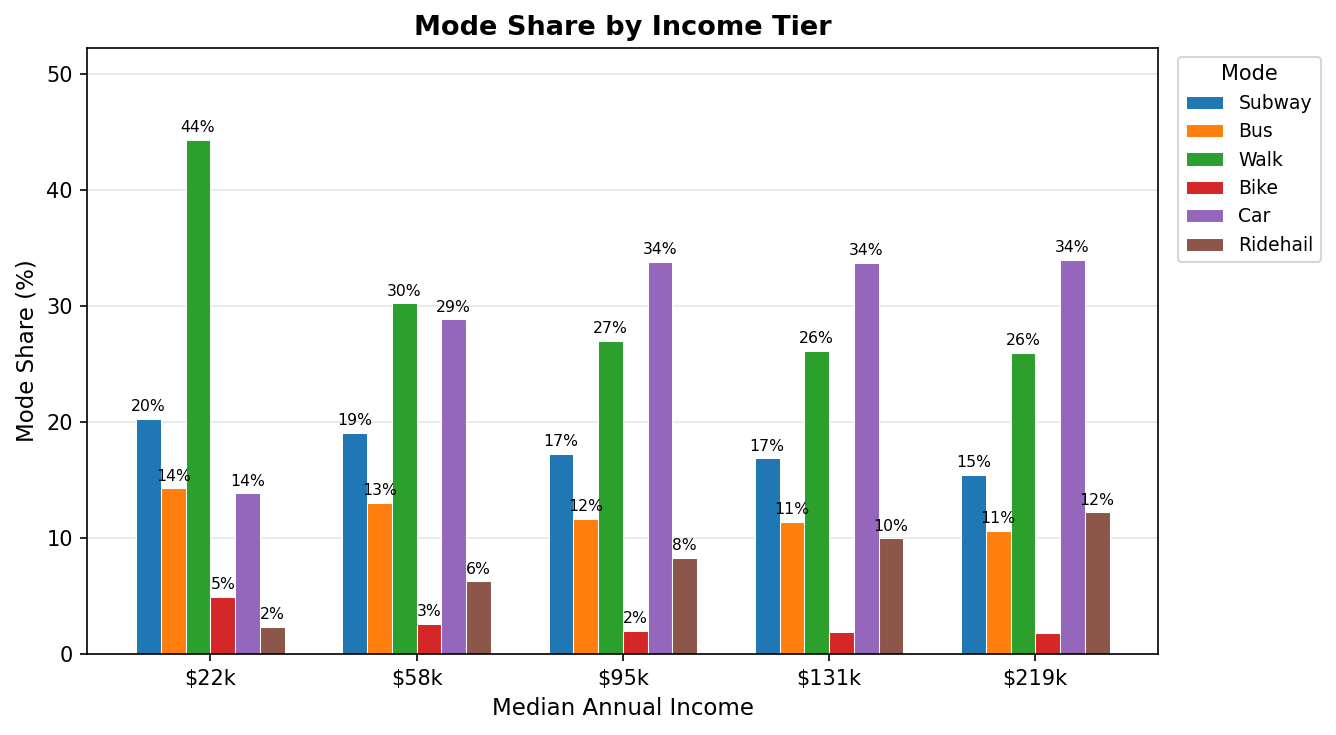}\vspace{-8pt}
        \caption{Baseline}
        \label{fig:mode_share_income_baseline}
    \end{subfigure}
    \begin{subfigure}{0.49\textwidth}
        \centering
        \includegraphics[width=0.95\textwidth]{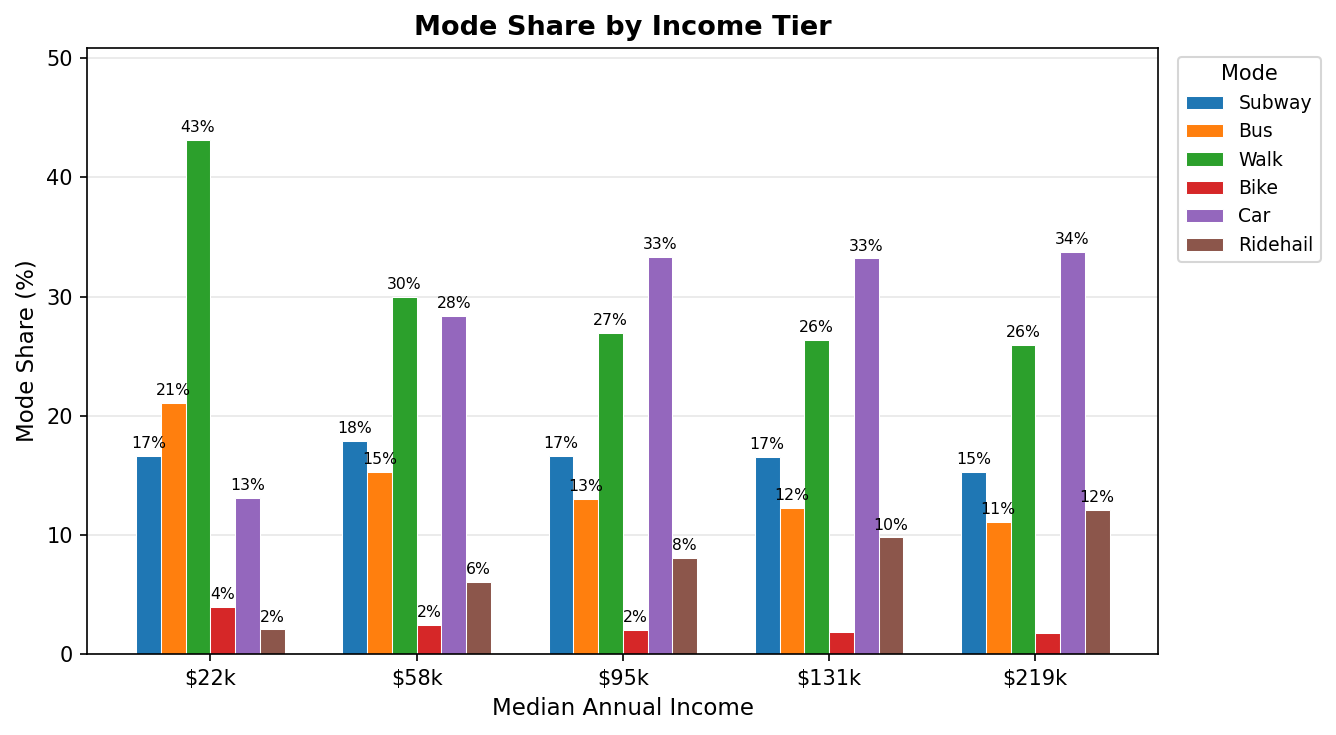}\vspace{-7pt}
        \caption{Scenario C}
        \label{fig:mode_share_income_scenario_c}
    \end{subfigure}
    \vspace{-8pt}\caption{Mode share by income tier, showing differences in transportation preferences across groups.}\vspace{-6pt}
    \label{fig:mode_share_income}
\end{figure*}

\subsubsection{OD Flow Validation}

Figure~\ref{fig:od_comparison} compares borough-level subway OD flows between observed MTA data and simulation outputs. Figure~\ref{fig:od_comparison_mta} shows observed average daily ridership and Figure~\ref{fig:od_comparison_baseline} presents simulated flows (averaged over two weeks).
%
Overall, the simulation captures the broad spatial structure of subway travel demand. In both observed and simulated matrices, major flows are concentrated within and between Manhattan and the outer boroughs, with particularly high volumes for intra-Manhattan trips and flows between Manhattan and Brooklyn or Queens. This indicates that the model reproduces the dominant travel corridors and directional patterns across the network.

However, the results reveal systematic deviations in magnitude. The simulation consistently overestimates ridership across most borough pairs, with ratio values exceeding 1.0 for nearly all flows. Overestimation is especially pronounced for intra-borough trips, such as within Brooklyn and Queens, where simulated flows exceed observed values by factors of 6.4 and 7.4, respectively. Similarly, Bronx–Bronx trips show the largest discrepancy, likely reflecting overestimation of short-distance trips in the model. 

Inter-borough flows involving Manhattan are also overestimated, though to a lesser extent. For example, Manhattan-to-Manhattan flows are overpredicted by a factor of about 2.45, while flows between Manhattan and other boroughs typically range between 1.3 and 3.0. These patterns suggest that while relative spatial distributions are well captured, the model tends to inflate absolute demand levels, particularly for short-distance or localized trips.
Low-volume flows, such as those involving Staten Island, exhibit greater variability and are less reliably estimated due to smaller baseline ridership levels and limited transit connectivity.

Overall, the OD comparison demonstrates that the simulation reproduces key structural patterns of travel demand but exhibits systematic upward bias in magnitude. These discrepancies are consistent with earlier observations and highlight the importance of calibration and scaling when translating activity-based trip generation into aggregate transit ridership. Despite these differences, the model provides a reasonable approximation of spatial flow patterns for comparing relative impacts of fare policies across boroughs.

\vspace{-5pt}
\subsubsection{Mode Share by Income Tier}

Figure~\ref{fig:mode_share_income_baseline} shows how modal choices vary across income groups, highlighting systematic differences in transportation behavior. Clear gradients emerge across modes as income increases.

Walking remains the dominant mode for the lowest-income group, accounting for 44\% of trips, but declines steadily with income to approximately 26\% among middle-income agents. In contrast, car usage exhibits the opposite pattern, increasing sharply from 14\% for the lowest-income agents to about 34\% for middle- and high-income groups, where it becomes the most common mode. This transition reflects increased access to private vehicles and reduced sensitivity to monetary travel costs at higher income levels.

Public transit modes show moderate variation across income tiers. Subway usage decreases gradually from 20\% to 15\%, while bus usage declines from 14\% to 11\%. These trends suggest that lower-income agents rely more heavily on transit, consistent with higher cost sensitivity and lower vehicle ownership.

Ridehail usage increases steadily with income, rising from 2\% among low-income agents to 12\% for high-income groups, reflecting greater willingness and ability to pay for convenience. Biking remains a minor mode across all groups, with a slight decline from 5\% to around 2\%.

Overall, the results highlight a strong income-dependent shift from active and transit modes toward private and on-demand modes. These patterns validate the role of income-normalized cost in the model, which produces heterogeneous fare sensitivities and mode substitution behavior across demographic groups.

\begin{figure*}
    \centering
    \includegraphics[width=0.9\linewidth]{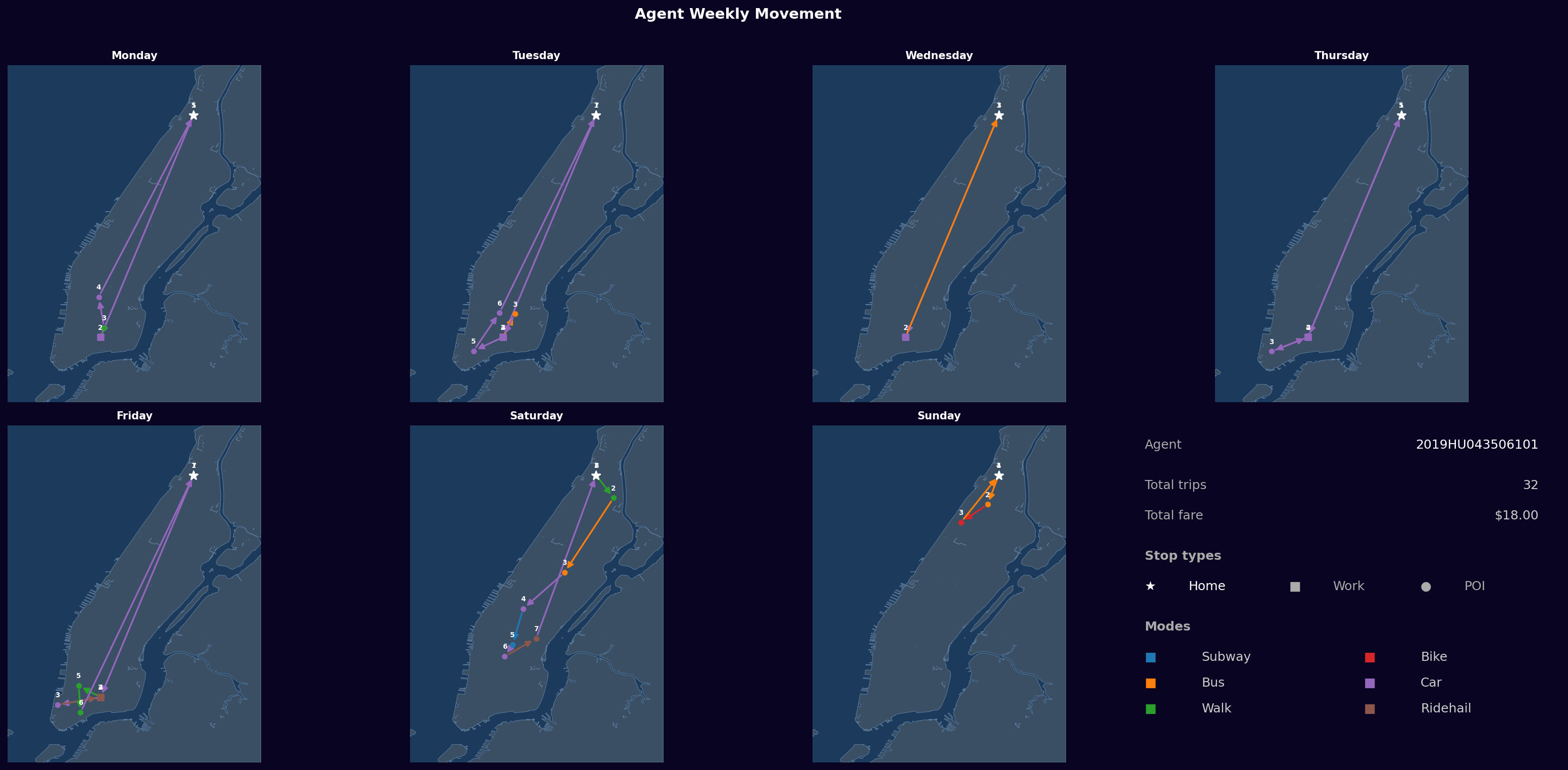}\vspace{-8pt}
    \caption{Weekly mobility trajectory of a representative agent, showing weekday commuting and weekend activity patterns.\vspace{-8pt}}
    \label{fig:agent_week}
\end{figure*}

\subsubsection{Individual Mobility Trajectories}

Figure~\ref{fig:agent_week} illustrates the weekly movement patterns of a representative agent, highlighting how daily activity schedules translate into spatial mobility trajectories. The figure shows consecutive daily paths across NYC, with locations categorized as home, work, and other activity points, and travel segments colored by mode.

Clear temporal structure is observed in weekday behavior. From Monday to Friday, the agent follows a regular commute pattern between a residential location in Lower Manhattan and a workplace in Upper Manhattan. These weekday trips are primarily direct and repeated, reflecting structured work-related travel. Minor variations are visible on certain days (e.g., Tuesday and Friday), where additional intermediate stops introduce short detours, indicating discretionary activities such as errands or social visits.

Weekend behavior contrasts sharply with weekday patterns. On Saturday, the trajectory becomes more complex, including multiple intermediate destinations distributed across different parts of Manhattan, suggesting flexible, multi-purpose activity chains. Sunday exhibits simpler, shorter trips concentrated near the home location, with reduced travel distances and fewer stops.

Mode usage also varies across the week. Commuter trips primarily rely on transit, while optional and weekend trips include a mix of modes, including walking and occasional alternative modes, reflecting context-dependent travel decisions.

Overall, this example demonstrates how the model captures both regular commuting behavior and flexible activity patterns within a unified framework. It highlights the ability of the simulation to represent realistic temporal variability, spatial movement, and multimodal travel decisions at the individual level.

\vspace{-6pt}
\subsection{Policy Impact Analysis}

We evaluate how different fare policies influence system-level outcomes.


\vspace{-5pt}
\subsubsection{Ridership}
Table~\ref{tab:ridership_results} reports total transit ridership under each policy scenario. Relative to the baseline (12.78M trips), Scenario~A yields a modest increase of 1.32\%, while Scenario~B results in a slight decline of 1.46\%. Scenario~C produces the largest increase, with ridership rising by 4.34\%.

\begin{table}[h]
\centering
\vspace{-3pt}\caption{Transit ridership for different fare policy scenarios.}\vspace{-8pt}
\label{tab:ridership_results}
\small
\begin{tabular}{lccc}
\toprule
\textbf{Scenario} & \textbf{Total Trips} & \textbf{Change (\%)} & \textbf{Subway / Bus Split} \\
\midrule
Baseline & 12,778,111 & -- & 7,569,636 / 5,208,475 \\
Scenario A & 12,946,944 & +1.32\% & 7,696,723 / 5,250,221 \\
Scenario B & 12,591,573 & -1.46\% & 7,451,683 / 5,139,890	 \\
Scenario C & 13,332,469 & +4.34\% & 7,001,757 / 6,330,712 \\
\bottomrule
\end{tabular}\vspace{-10pt}
\end{table}

Changes in modal composition reveal heterogeneous behavioral responses across income groups. Under Scenarios~A and B, subway ridership remains dominant with only minor adjustments in the subway-to-bus split. 
In contrast, Scenario~C leads to a substantial increase in bus usage, particularly among lower-income groups. As shown in Figure~\ref{fig:mode_share_income_scenario_c}, bus share rises to approximately 21\% for the lowest-income group, accompanied by a corresponding reduction in subway usage. This indicates a strong mode substitution effect within transit, driven by fare changes that disproportionately benefit bus travel.
For higher-income groups, the effect is more limited. Car usage remains the dominant mode (around 33–34\%), with relatively modest changes in transit shares. Walking also remains significant across all income groups but declines with income in a similar pattern to the baseline.

Overall, Scenario~C induces a redistribution within transit modes rather than a uniform shift across all groups, with the largest behavioral response concentrated among lower-income travelers. This highlights the income-sensitive impact of fare policy on mode choice. Figure~\ref{fig:mode_share_income_scenario_c} provides a detailed breakdown of these patterns.

These results address RQ1 by quantifying how alternative fare policies influence both overall ridership and modal distribution. While total demand responds moderately to pricing changes, the redistribution between transit modes highlights the sensitivity of traveler decisions to relative costs across modes.

\vspace{-5pt}
\subsubsection{Revenue}
Table~\ref{tab:revenue_results} summarizes fare revenue impacts. While fare increases decrease ridership, they result in a 14.66\% increase in revenue in Scenario B, highlighting a potential trade-off between accessibility and financial sustainability.

\begin{table}[h]
\centering
\vspace{-5pt}\caption{Fare revenue comparison across policy scenarios.}\vspace{-6pt}
\label{tab:revenue_results}
\small
\begin{tabular}{lccc}
\toprule
\textbf{Scenario} & \textbf{Total Revenue (\$)} & \textbf{Change (\%)} & \textbf{Revenue per Trip} \\
\midrule
Baseline & 38,334,333.00 & -- & 3.00 \\
Scenario A & 40,740,227.61 & +6.28\% & 3.15 \\
Scenario B & 44,070,506.00 & +14.96\% & 3.50 \\
Scenario C & 21,005,271.00 & -45.21\% & 1.54 \\
\bottomrule
\end{tabular}\vspace{-10pt}
\end{table}

These findings address RQ2 by illustrating the trade-off between increased ridership and reduced fare revenue under pricing interventions. This trade-off has direct implications for system-wide financial sustainability.

\vspace{-5pt}
\subsubsection{Equity}
\begin{figure}
    \begin{subfigure}{0.49\textwidth}\vspace{-5pt}
        \centering
        \includegraphics[width=0.75\textwidth]{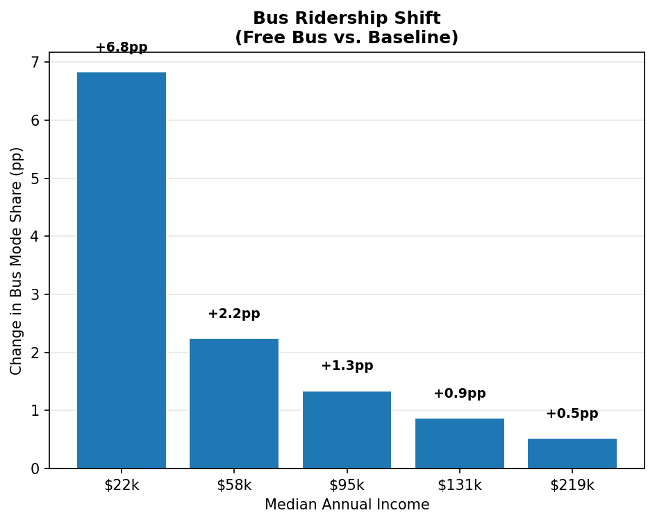}\vspace{-7pt}
        \caption{Equity Ridership Shift}
        \label{fig:equity_results_ridership}
    \end{subfigure}
    \begin{subfigure}{0.49\textwidth}
        \centering
        \includegraphics[width=.75\textwidth]{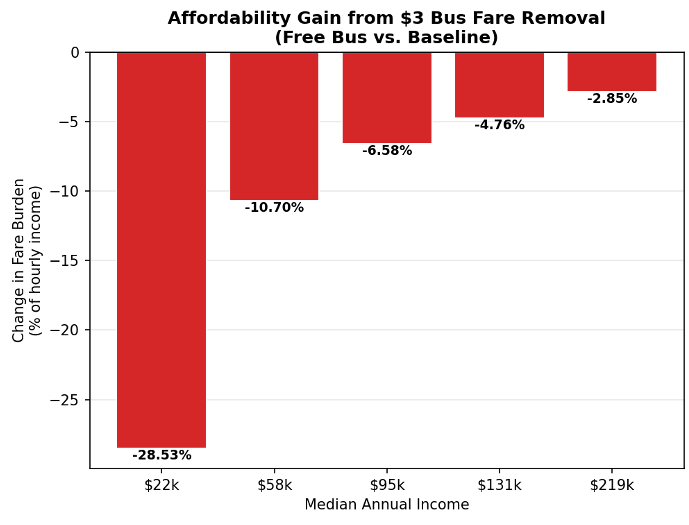}\vspace{-7pt}
        \caption{Affordability Gain}\vspace{-10pt}
        \label{fig:equity_results_affordability}
    \end{subfigure}
    \caption{Equity impacts of fare-free buses. Shows measurable improvements in accessibility for low-income residents compared to baseline}\vspace{-14pt}
    \label{fig:equity_results}
\end{figure}

To evaluate RQ3, we measure equity using change in bus ridership and affordability gain from Scenario C, fare-free buses. Figure~\ref{fig:equity_results} shows that eliminating the \$3 bus fare delivers a clearly progressive benefit. Figure~\ref{fig:equity_results_ridership} illustrates shifts in bus mode share. Low-income riders exhibit the largest increase, with bus share rising by approximately +6.8 percentage points for the lowest-income group (\$22k), compared to only +0.5 percentage points for the highest-income group (\$219k). This indicates that fare elimination induces substantially greater mode switching toward buses among lower-income populations.
Figure~\ref{fig:equity_results_affordability} shows affordability gains measured as reductions in fare burden (relative to income). The benefits are highly unequal across groups: the lowest-income group experiences a reduction of approximately 28.5\% in fare burden, while the highest-income group sees a much smaller decrease of about 2.9\%. These results demonstrate that a flat fare removal yields disproportionately larger financial relief for riders with lower incomes.

These results show that fare policies can significantly alter accessibility across income groups, directly addressing RQ3. In particular, policies such as Scenario B reduce disparities by improving transit access for lower-income populations.

\vspace{-8pt}
\subsection{Sampling Efficiency}
\label{sec:sampling_eff}
\begin{figure}[t]
  \centering
  \includegraphics[width=1\linewidth]{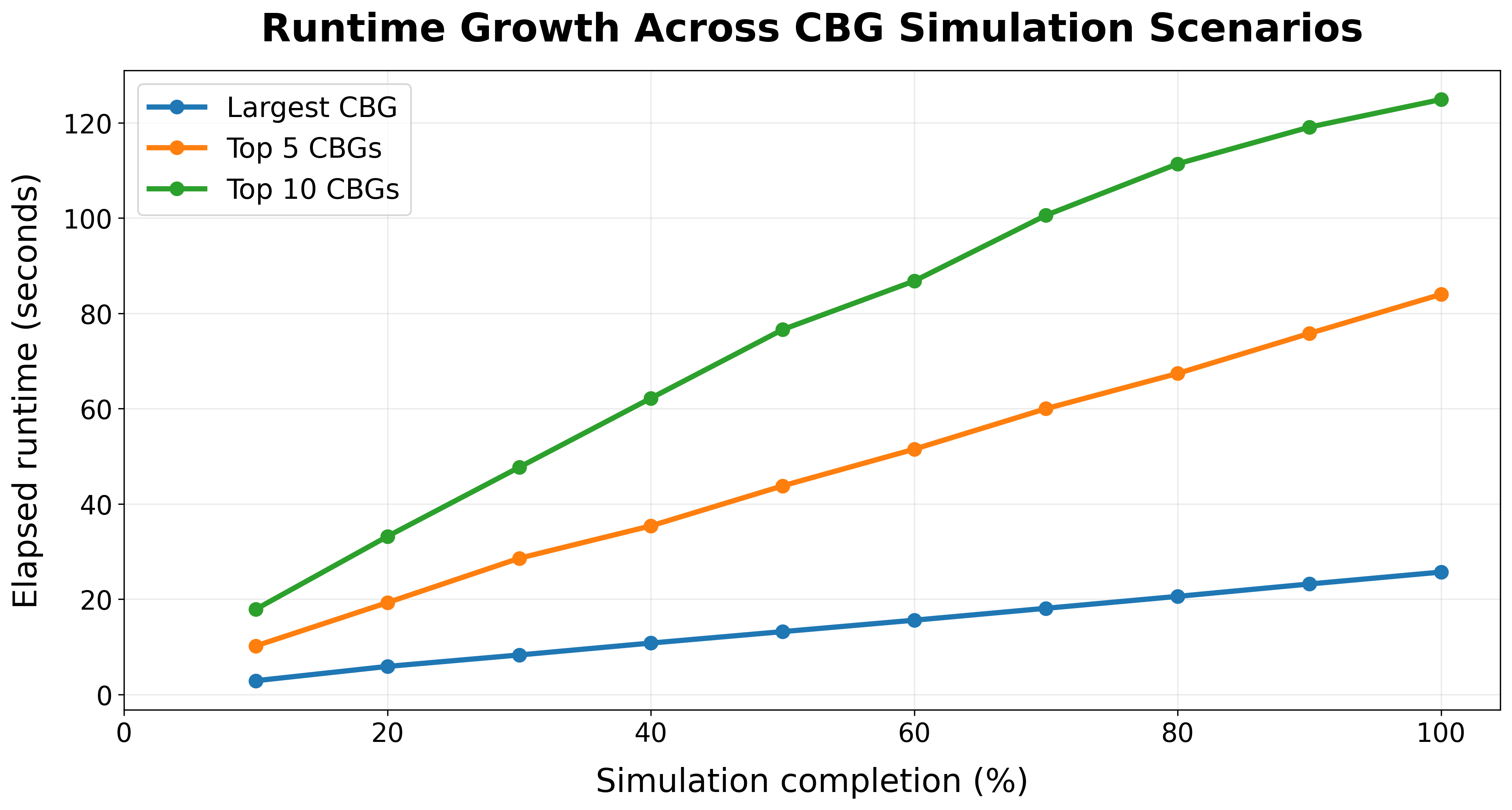}\vspace{-5pt}
  \caption{Elapsed simulation runtime for the largest, top-5, and top-10 CBG groups by agent count.}\vspace{-10pt}
  \label{fig:load_test}
\end{figure}

This section evaluates whether the proposed sampling framework can reduce computational cost while preserving benchmark-level aggregate patterns. To characterize runtime behavior, a load test measures elapsed simulation time as a function of the proportion of agents processed. Figure~\ref{fig:load_test} shows the results for the largest CBG by agent count, as well as aggregated runs for the top-5 and top-10 largest CBGs. Runtime increases approximately linearly with the number of agents processed, suggesting that simulation cost is primarily driven by agent count rather than nonlinear bottlenecks.

Runtime measurements were collected on a workstation running Ubuntu 24.04.4 LTS with an Intel Core Ultra 9 285K processor, 24 physical cores, and 62.2\,GB of RAM. Simulations used Python~3.11.15 in a single-process CPU setting without GPU acceleration. 
The largest CBG contains 2,424 agents and completes in 25.7 seconds, while the top-5 and top-10 aggregated runs require 84.0 and 124.9 seconds, respectively. Since agent counts vary substantially across the 6,395 CBGs, with a long-tailed distribution \ifthenelse{\boolean{arxiv}}{
(cf. Figure~\ref{fig:num_agent_cbg} in Appendix~\ref{sec:appendix_sampling}).
}
{
\cite{wischhover26simulating_arxiv}.
}
Using the largest-CBG runtime as a conservative per-CBG estimate, a sequential simulation over all 6,395 CBGs would require approximately $6{,}395 \times 25.7 \approx 164{,}000$ seconds, or roughly 45 hours for a single policy evaluation. Multiple policy scenarios or repeated trials would increase this cost proportionally. With $k=6$ clusters and 10 representative CBGs per cluster, the sampling framework simulates 60 CBGs, corresponding to approximately 26 minutes under the same conservative runtime assumption.

\begin{figure}[t]
    \centering
    \includegraphics[width=0.98\linewidth]{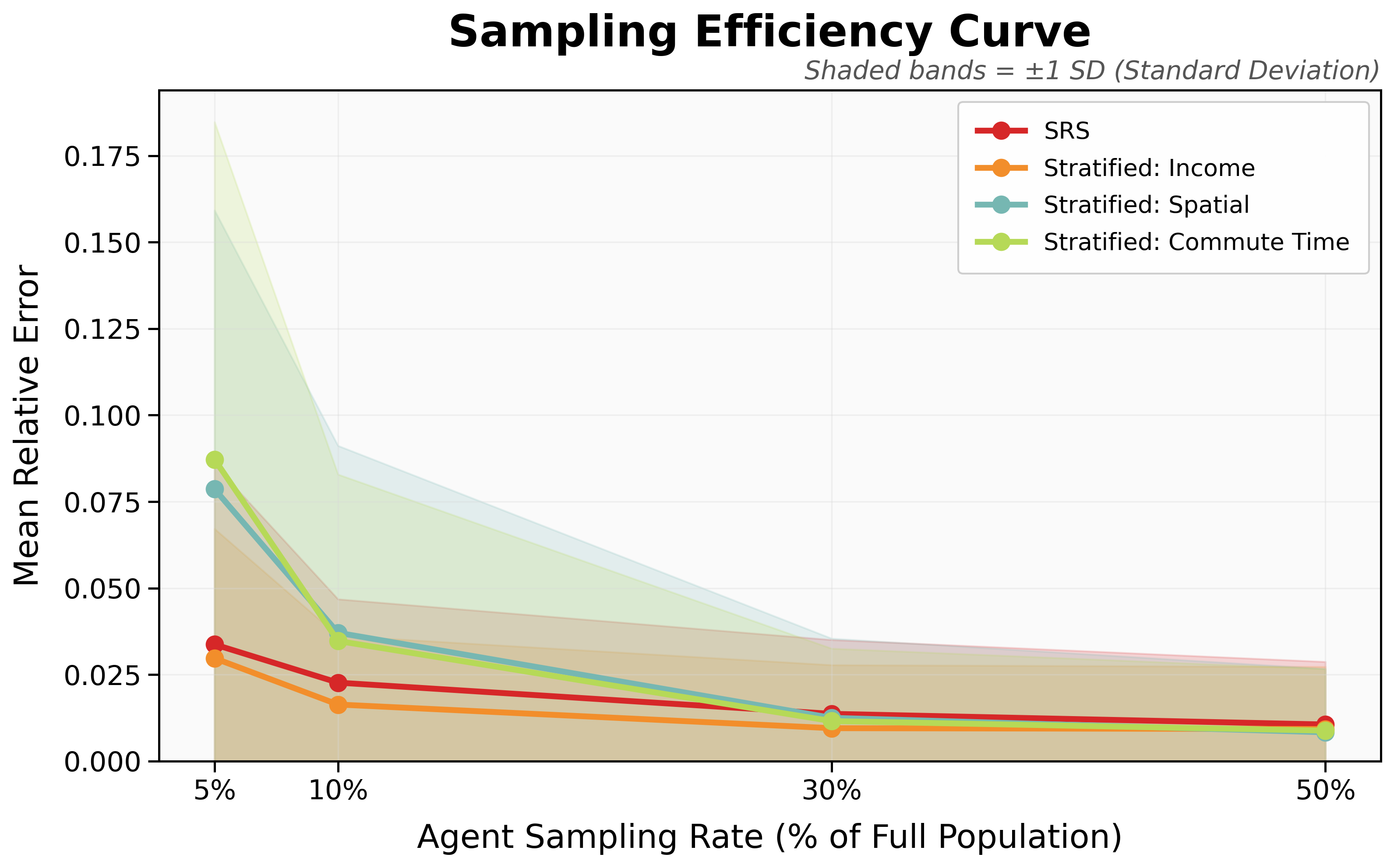}\vspace{-5pt}
    \caption{ Sampling efficiency by agent sampling rate. Shaded bands denote $\pm 1$ standard deviation across 20 repeated simulations.}\vspace{-10pt}
    \label{fig:sampling_efficiency_curve}
\end{figure}

Figure~\ref{fig:sampling_efficiency_curve} reports sampling accuracy from 20 repeated simulations for each strategy and sampling rate. Mean relative error is computed against the 100\% benchmark simulation and averaged over the target outcome metrics. Error generally decreases as the agent sampling rate increases.
Income-based stratification shows the lowest mean error in this experiment, although differences across strategies become smaller at higher sampling rates. Wider uncertainty bands at low sampling rates indicate greater variation across repeated samples.

Overall, these results suggest that representative CBG selection and stratified within-CBG agent sampling can reduce simulation cost while maintaining close agreement with the benchmark aggregate outputs in this experiment.

%% file: sections/conclusions.tex
\vspace{-5pt}
\section{Conclusion}
\label{sec:conclusion}

This paper presents a scalable, data-driven simulation framework for evaluating public transit fare policies in NYC. By integrating synthetic population modeling, agent-based behavioral simulation, multimodal travel-time estimation, and fare-sensitive mode choice, the framework enables detailed assessment of policy impacts on ridership, revenue, and equity.

Experimental results demonstrate that the framework reproduces realistic urban mobility patterns and supports systematic evaluation of alternative fare scenarios. The analysis reveals important trade-offs between accessibility, financial sustainability, and equity. In particular, fare policies can significantly influence travel behavior while generating heterogeneous impacts across income groups, highlighting the importance of equity-aware policy design.
To enable city-scale analysis, we introduce a sampling-based strategy that substantially reduces computational cost while preserving aggregate accuracy. This allows efficient exploration of multiple policy scenarios without requiring full-population simulation, making the framework practical for iterative decision support.

Future work will incorporate additional behavioral factors such as service reliability, crowding, and real-time adaptation. We also plan to extend the framework to model dynamic demand–supply interactions and learning-based pricing strategies, as well as apply it to other cities to assess generalizability.

%% file: sections/calibration_parameters.tex
\section{Parameter Calibration Details}
\label{sec:appendix_calibration}

\begin{table*}[t]
\centering
\caption{Calibration parameters and search ranges}\vspace{-6pt}
\small
\begin{tabular}{p{3cm} p{8.5cm} p{5.2cm}}
\hline
\textbf{Category} & \textbf{Parameters} & \textbf{Search Range} \\
\hline

Destination Scoring 
& \texttt{BETA\_BASE}, \texttt{V\_J\_ALPHA} 
& $(0.0, 5.0)$; $(0.0, 1.0)$ \\

Worker Schedule 
& \texttt{LUNCH\_PROB}, \texttt{DISCRETIONARY\_PROB}, \texttt{POST\_WORK\_STOPS\_P} 
& $(0.0, 1.0)$; $(0.0, 1.0)$; $[p_1, p_2],\ p_1+p_2=1$ \\

Non-worker Activity 
& \texttt{GO\_HOME\_BASE\_PROB}, \texttt{OUTING\_STOPS\_LAMBDA}, \texttt{OUTING\_STOPS\_MAX} 
& $(0.0, 1.0)$; $(0.1, 50.0)$; $(0, 50)$ \\

Home Dwell Time 
& \texttt{HOME\_DWELL\_MEAN\_MIN}, \texttt{HOME\_DWELL\_SD\_MIN}, \texttt{HOME\_DWELL\_BOUNDS} 
& $(10, 500)$; $(10, 500)$; $[(30,120),(120,480)]$ \\

Travel Cost 
& \texttt{BUS\_PENALTY}, \texttt{BUS\_MAX\_CAR\_RATIO}, \texttt{MAX\_TRAVEL\_MIN} 
& $(0.0, 50.0)$; $(0.0, 50.0)$; $(30, 180)$ \\

Fare Sensitivity 
& \texttt{FARE\_SENSITIVITY}, \texttt{COST\_BURDEN\_THRESHOLD}, \newline \texttt{COMMUTE\_FARE\_DAMPING}, \texttt{FARE\_DEST\_GAMMA} 
& $(0.0, 50.0)$; $(0.0, 10.0)$; \newline $(0.0, 10.0)$; $(0.0, 50.0)$ \\

Temporal Scheduling 
& \texttt{DEPART\_MEAN}, \texttt{DEPART\_SD}, \texttt{DEPART\_BOUNDS} 
& $[0,23]$; $(0.1, 5.0)$; $[(0.1,20.0),(20.0,500.0)]$ \\

& \texttt{RETURN\_MEAN}, \texttt{RETURN\_SD}, \texttt{RETURN\_BOUNDS} 
& $[0,23]$; $(0.1, 5.0)$; $[(0.1,20.0),(20.0,500.0)]$ \\

\hline
\end{tabular}\vspace{-6pt}
\label{tab:calibration}
\end{table*}

This appendix provides detailed information on the parameter space and calibration procedure. While the main text summarizes the optimization framework, this section documents the parameterization and constraints explored during calibration.

\vspace{-6pt}
\subsection{Parameterization}

The model includes parameters governing destination choice, activity scheduling, travel cost perception, and fare sensitivity. These parameters jointly determine how agents generate trips and respond to network and pricing conditions.

Table~\ref{tab:calibration} summarizes all calibrated parameters, grouped into behaviorally meaningful categories: (i) destination scoring, (ii) activity scheduling, (iii) home dwell behavior, (iv) travel cost perception, (v) fare sensitivity, and (vi) temporal scheduling.

\vspace{-6pt}
\subsection{Search Space Design}

Parameter ranges are intentionally broad to capture uncertainty in behavioral assumptions while enabling exploration of diverse mobility patterns. Continuous parameters are sampled within bounded intervals, while structured parameters (e.g., probability vectors and temporal bounds) are subject to additional constraints. For example, probability parameters are normalized, and temporal bounds must satisfy valid ordering constraints.

These design choices balance flexibility and tractability. Broad ranges support behavioral realism, while structural constraints improve efficiency during optimization.

\vspace{-6pt}
\subsection{Discussion}

The calibration framework preserves interpretability by associating each parameter with a clear behavioral meaning. This facilitates analysis of model sensitivity and policy impacts, particularly for fare-related interventions. The calibrated parameters provide a consistent foundation for evaluating transit fare scenarios in NYC.

%% file: sections/sampling_detail.tex
\vspace{-6pt}
\section{Sampling Method Details}
\label{sec:appendix_sampling}

This section provides detailed descriptions of the sampling framework used to reduce computational cost while preserving accuracy.

\vspace{-6pt}
\subsection{Sampling Framework Overview}

To reduce computational cost while preserving accuracy, we adopt a two-stage sampling framework. Running the full synthetic population of \num{3409887} agent rows for every candidate fare policy is computationally expensive. We therefore use a two-stage experimental design. First, CBGs are described by aggregate population features, standardized, and clustered; representative CBGs are then selected from each cluster for controlled experiments. Second, agents are sampled within each selected CBG using simple random or stratified designs, and the resulting simulation outputs are compared with a full-population benchmark. The first-stage CBG selection is used only to reduce the cost of the validation experiment. The final sampling rule is intended to be applied within every CBG.

\vspace{-6pt}
\subsection{CBG-Level Feature Construction}
\label{sec:cbg_feature_construction_app}

For each of the \num{6395} home CBGs, agent-level attributes were aggregated into the 23 baseline features listed in Table~\ref{tab:cbg_features_exp3}. Each row in the input population was treated as a simulated agent. Therefore, \texttt{represented\_\allowbreak population} denotes the number of agent rows in a CBG, and \texttt{represented\_\allowbreak population\_\allowbreak log} is calculated as $\log(1+N_c)$, where $N_c$ is the number of rows in CBG $c$. 

The log transformation retains information about relative population size while reducing the influence of very large CBGs, preventing clustering from being dominated by population differences. Raw centroid coordinates were excluded from the baseline feature set to avoid clusters being driven primarily by geographic proximity rather than socioeconomic, mobility, and behavioral characteristics.

Categorical diversity is measured using normalized Shannon entropy, capturing how evenly categories are distributed within each CBG. The entropy is normalized to the range $[0,1]$ using only observed categories. This measure is applied to variables such as income tier, vehicle ownership, age group, commute mode, and work-destination group. Ages are grouped as under 18, 18--34, 35--64, and 65 or older. Home--work distance is calculated using haversine distance between centroids.

\begin{figure}[t]
  \centering
  \includegraphics[width=1\linewidth]{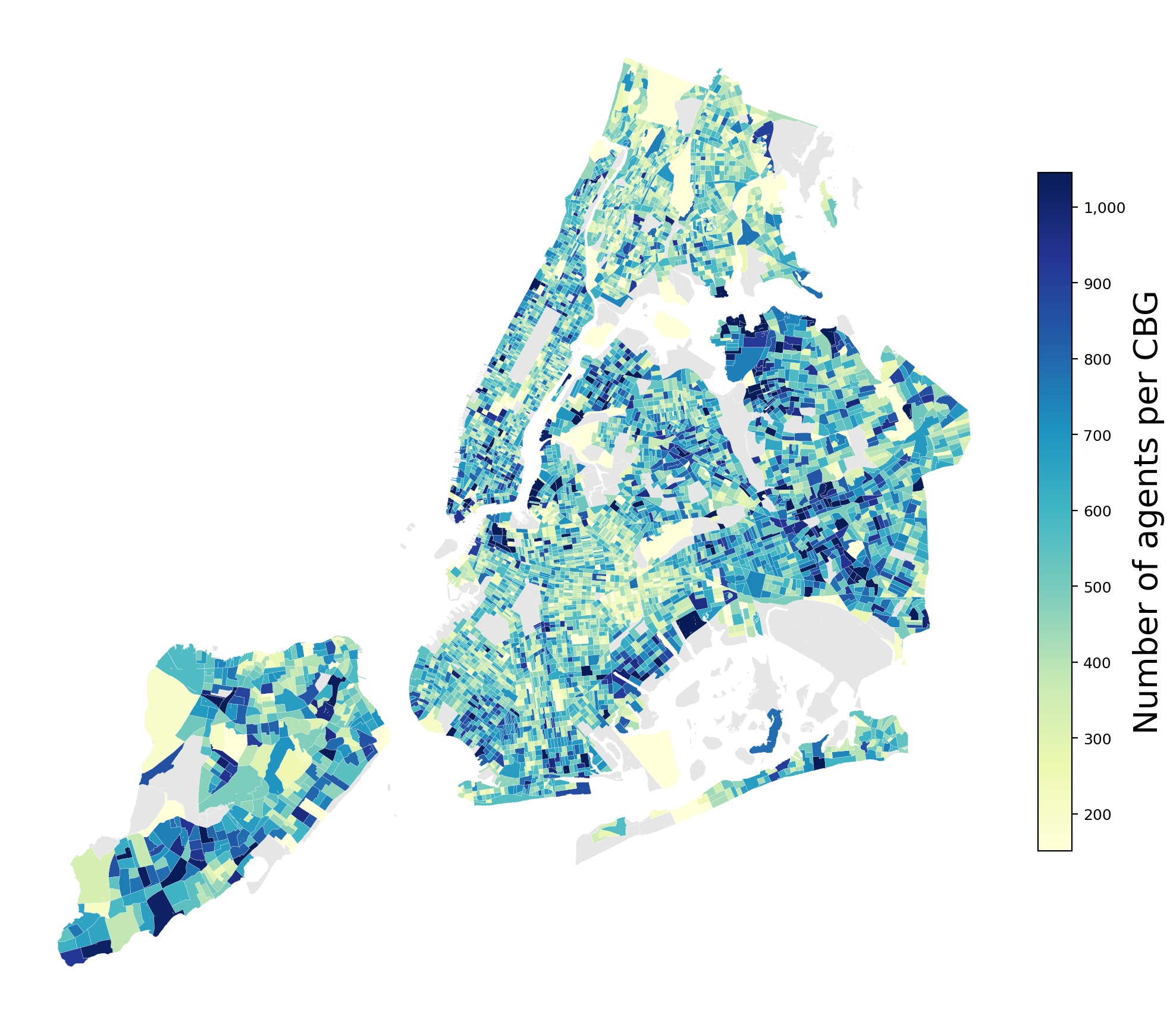}\vspace{-10pt}
  \caption{Distribution of agent counts across NYC CBGs. The distribution is highly skewed, with a small number of CBGs containing large agent populations, motivating the use of sampling.}\vspace{-6pt}
  \label{fig:num_agent_cbg}
\end{figure}

\begin{table}[h] 
\centering
\caption{Baseline CBG features used for clustering. All 23 features are standardized only after aggregation and immediately before clustering.}\vspace{-6pt}
\label{tab:cbg_features_exp3}
\small
\setlength{\tabcolsep}{3pt} 
\renewcommand{\arraystretch}{1.08}
\begin{tabular}{@{}ll@{}}
\toprule
\textbf{Category} & \textbf{CBG-level feature (aggregation)} \\
\midrule
Size
  & \texttt{represented\_population\_log} ($\log(1+N_c)$) \\
\midrule
Socioeconomic
  & \texttt{pincp\_median} (median personal income) \\
  & \texttt{hincp\_median} (median household income) \\
  & \texttt{income\_tier\_entropy} (normalized entropy) \\
  & \texttt{vehicle\_ownership\_share} (share with $\mathrm{veh}>0$) \\
  & \texttt{vehicle\_entropy} (normalized entropy) \\
  & \texttt{worker\_share} (NYC-worker share) \\
  & \texttt{age\_mean} (mean age) \\
  & \texttt{age\_entropy} (entropy of four age groups) \\
\midrule
Spatial interaction
  & \texttt{work\_cbg\_entropy} (work-destination entropy) \\
  & \texttt{top1\_work\_cbg\_share} (largest destination share) \\
  & \texttt{median\_home\_work\_distance\_km} (median) \\
  & \texttt{p90\_home\_work\_distance\_km} (90th percentile) \\
  & \texttt{mean\_commute\_minutes\_declared} (mean) \\
  & \texttt{poi\_candidate\_count\_log} ($\log(1+n_{\mathrm{POI}})$) \\
\midrule
Behavioral
  & \texttt{jwtrns\_entropy} (commute-mode-code entropy) \\
  & \texttt{pref\_income\_sensitivity\_mean} (mean) \\
  & \texttt{pref\_distance\_aversion\_mean} (mean) \\
  & \texttt{pref\_transit\_reliance\_mean} (mean) \\
  & \texttt{work\_arrive\_hour\_mean} (mean) \\
  & \texttt{work\_depart\_hour\_mean} (mean) \\
  & \texttt{lunch\_hour\_mean} (mean) \\
  & \texttt{schedule\_observed\_share} (complete schedule share) \\
\bottomrule
\end{tabular}\vspace{-10pt}
\end{table}

\vspace{-6pt}
\subsection{CBG Clustering and Representative Selection}
\label{sec:cbg_clustering_app}

Before clustering, missing values are imputed using feature-wise medians, and all features are standardized:
\begin{equation}
  z_{cj} = \frac{x_{cj} - \mu_j}{\sigma_j}.
\end{equation}

We apply $K$-Means clustering with $K \in [2,10]$. The final $K$ is selected using a composite score combining Silhouette, Calinski--Harabasz, Davies--Bouldin, and elbow criteria. We choose the smallest $K$ within 0.02 of the best score.

From each cluster, representative CBGs are selected to span both central and boundary regions. The number of representatives is 3--10 per cluster (approximately 1\% of eligible CBGs). Eligibility requires at least 10 agents per CBG.

\vspace{-6pt}
\subsection{Within-CBG Agent Sampling}
\label{sec:within_cbg_app}

For each representative CBG, we test sampling rates from 10\% to 90\% of agents. All sampling is without replacement. Table~\ref{tab:strategies} summarizes the strategies used.

\begin{table}[t]
\centering
\caption{Active within-CBG sampling strategies and stratum definitions.}\vspace{-5pt}
\label{tab:strategies}
\small
\setlength{\tabcolsep}{2pt}
\renewcommand{\arraystretch}{1.10}
\begin{tabular}{@{}lp{5.0cm}@{}}
\toprule
\textbf{Strategy} & \textbf{Within-CBG strata} \\
\midrule
\textsc{srs}                       & None; uniform simple random sampling \\
\textsc{stratified\_income}        & income tier \\
\textsc{stratified\_spatial}       & income tier $\times$ vehicle ownership
                                     $\times$ work-destination group \\
\textsc{stratified\_commute\_time} & income tier $\times$ commute-time bin \\
\bottomrule
\end{tabular}
\end{table}

For stratified designs, samples are allocated proportionally, with at least one agent per stratum. Sampling weights are computed as $w_i = \pi_i^{-1}$ to scale results to the full population.

\vspace{-6pt}
\subsection{Sampling Error Metrics}
\label{sec:SamplingErrorMetrics}

For each selected CBG \(c\), sampling strategy \(s\), sampling rate \(r\), and target metric \(k\), we compared the sampled simulation result against the corresponding repeated \(100\%\) population benchmark. Let \(\bar{Y}^{\mathrm{sample}}_{c,k}(r,s)\) denote the mean metric value from the sampled simulation condition, and let \(\bar{Y}^{100\%}_{c,k}\) denote the mean value from the full-population benchmark. For scalar outcome metrics, the relative error was computed as
\[
e_{c,k}(r,s)
=
\frac{
\left|
\bar{Y}^{\mathrm{sample}}_{c,k}(r,s)
-
\bar{Y}^{100\%}_{c,k}
\right|
}{
\left|
\bar{Y}^{100\%}_{c,k}
\right|
}.
\]

The sampling efficiency curve summarizes these errors by sampling rate and strategy. Specifically, the plotted value on the \(y\)-axis is the unweighted mean relative error across all selected CBGs and target metrics:
\[
\bar{e}(r,s)
=
\frac{1}{|\mathcal{C}|\,|\mathcal{K}|}
\sum_{c \in \mathcal{C}}
\sum_{k \in \mathcal{K}}
e_{c,k}(r,s),
\]
where \(\mathcal{C}\) is the set of selected CBGs and \(\mathcal{K}\) is the set of target evaluation metrics. The shaded band represents \(\pm 1\) standard deviation of the CBG--metric-level error values used to compute \(\bar{e}(r,s)\).

Mode-share error was computed separately because it compares a distribution rather than a scalar outcome. For each CBG and sampling condition, we computed the RMSE between the sampled and benchmark mode-share vectors:
\[
\mathrm{RMSE}_{c}(r,s)
=
\sqrt{
\frac{1}{M}
\sum_{m=1}^{M}
\left(
p^{\mathrm{sample}}_{c,m}(r,s)
-
p^{100\%}_{c,m}
\right)^2
},
\]
where \(p^{\mathrm{sample}}_{c,m}(r,s)\) and \(p^{100\%}_{c,m}\) are the sampled and benchmark shares for travel mode \(m\). This RMSE was included as an additional target metric in the aggregate sampling efficiency curve.